\documentclass[journal,twoside]{IEEEtran}
\usepackage{amsmath,amssymb,amsfonts}
\usepackage{graphicx}
\usepackage[caption=false,font=normalsize,labelfont=rm,textfont=rm]{subfig}
\usepackage{hyperref}

\begin{document}
\bstctlcite{IEEEexample:BSTcontrol}

\title{The Expected Peak-to-Average Power Ratio of White Gaussian Noise in Sampled I/Q Data
\author{Adam Wunderlich and Aric Sanders}
\thanks{
The authors are with the Communications Technology Laboratory, National Institute of Standards and Technology, Boulder, CO 80305 USA. (e-mail: adam.wunderlich@nist.gov, aric.sanders@nist.gov). 

Digital Object Identifier 10.1109/TIM.2025.3544288
}}

\IEEEpubid{U.S. Government work not protected by U.S. copyright.}

\maketitle

\begin{abstract}
One of the fundamental endeavors in radio frequency (RF) metrology is to measure the power of signals, where a common aim is to estimate the peak-to-average power ratio (PAPR), which quantifies the ratio of the maximum (peak) to the mean value.  For a finite number of discrete-time samples of baseband in-phase and quadrature (I/Q) white Gaussian noise (WGN) that are independent and identically distributed with zero mean, we derive a closed-form, exact formula for mean PAPR that is well-approximated by the natural logarithm of the number of samples plus Euler’s constant.  Additionally, we give related theoretical results for the mean crest factor (CF).  After comparing our main result to previously published approximate formulas, we examine how violations of the WGN assumptions in sampled I/Q data result in deviations from the expected value of PAPR.  Finally, utilizing a measured RF I/Q acquisition, we illustrate how our formula for mean PAPR can be applied to spectral analysis with spectrograms to verify when measured RF emissions are WGN in a given frequency band.       
\end{abstract}


\section{Introduction}

\IEEEPARstart{I}{n radio} frequency (RF) electronic measurements, it is common to perform peak detection, where the maximum amplitude or power of data collected over a given time interval is recorded.  In particular, the peak-to-average power ratio (PAPR) and crest factor (CF), equal to the square root of PAPR, arise in the design of communication signals and power amplifiers for RF transmitters \cite{Rahmatallah2013, Nikandish2020, Dunsmore2020}.  While it is important to understand the PAPR and CF for transmitted continuous-time communication signals and their impact on power amplifiers, it is equally important to characterize PAPR and CF for sampled RF measurements of received signals, e.g., with a signal analyzer.  Notable applications of received signals include spectrum monitoring \cite{Cotton2014, Kuester2022} and the detection of transient electromagnetic interference (EMI) \cite{Stienne2020}.  The appropriate selection of settings for measurement instrumentation in these domains requires foreknowledge of the peak level that can be expected over a given time period. 

Modulated RF signals are typically measured, processed, and stored in an equivalent baseband representation, called the in-phase and quadrature (I/Q) signal representation \cite{Proakis2002}.  Furthermore, the noise level in the absence of a signal is typically dominated by thermal noise in the in-phase (I) and quadrature (Q) components that can be modeled as a stationary, zero-mean, white Gaussian process \cite{Vasilescu2006, Milotti2019}. Therefore, the expected PAPR of stationary white Gaussian noise (WGN) in I/Q data is of high interest to differentiate thermal noise from other noise and signal types.

The difference between PAPR assessment for continuous-time and discrete-time measurements is often ignored \cite{Wulich2000}, because the Nyquist-Shannon sampling theorem implies that a band-limited signal may be perfectly reconstructed from samples if the sampling rate is greater than or equal to the Nyquist rate.  However, real signals are not perfectly band-limited, and the analog-to-digital (A/D) and digital-to-analog (D/A) conversion process is non-ideal \cite{Mitra2006}.  Consequently, the PAPR of a discrete-time sample is not identical to the PAPR for the corresponding continuous-time signal.  Bounds on the difference between the peak of a continuous-time, band-limited signal and the peak value of a discrete-time, over-sampled measurement were studied in \cite{Wunder2003} and \cite{Boche2020}.

Extrema theory for continuous-time, stationary random processes is reviewed in \cite{Leadbetter1983} and \cite{Wirsching1995}. In particular, notable results concerning the expected number of local peaks above a specified level per unit time were developed for band-limited, continuous-time signals in \cite{Rice1944, Cartwright1956}.  This theory was later applied by Ochai and Imai \cite{Ochiai2001} to derive accurate expressions for the PAPR distribution of band-limited, continuous-time orthogonal frequency-division multiplexing (OFDM) communication signals.  

\IEEEpubidadjcol 

The present work concerns the expected PAPR for discrete-time, sampled I/Q measurements of WGN, which is relevant to digital signal analyzer measurements. Namely, using extreme value theory \cite{Castillo1988} and the theory of order statistics \cite{David1981}, we develop highly accurate, closed-form formulas for the mean PAPR and mean CF for a finite collection of $n$ independent, identically distributed (i.i.d.) I/Q samples of stationary, zero-mean WGN.  To our knowledge, the theoretical results given here and their connection to measurement have not been presented in the literature. For this problem, approximate formulas for the mean PAPR were presented in \cite{Dunsmore2020} and \cite{Keysight2021}.  Also, exact cumulative distribution functions (CDFs) of PAPR and CF for WGN I/Q samples were given in \cite{Ochiai2001}, but not analyzed further.  Compared to these prior works, we present additional theoretical insights, context, and applications.  Namely, we examine how non-ideal I/Q measurements of thermal noise can violate the WGN assumptions and impact the expected PAPR.  Further, we show how the expected mean PAPR formula can be applied to spectrograms, which are widely used to visualize the frequency content of a nonstationary signal as it varies in time, to verify when measured RF emissions in a given frequency band consist of thermal noise.

\section{Theoretical Results}
\label{sec:theory_section}

Consider a baseband signal, $X$, with I/Q signal components, $X_I$ and $X_Q$.  Denote the amplitude (or envelope) of $X$ as $A = \sqrt{|X_I|^2 + |X_Q|^2}$ and the instantaneous power (or squared amplitude)\footnote{For generality, we avoid assuming physical units and the requisite scale factors necessary to convert between squared amplitude and power units.} as $P = |X_I|^2 + |X_Q|^2$.  Unless otherwise stated, we assume that the I/Q components can be modeled as WGN.  Specifically, $X_I$ and $X_Q$, are taken to be i.i.d. normal random variables with zero mean and variance $\sigma^2$, denoted $X_I \sim \mathcal{N}(0,\sigma^2)$ and $X_Q \sim \mathcal{N}(0,\sigma^2)$. 

We denote sample numbers with subscripts, and order statistics with subscripts in parentheses.  For example, given a sample of $n$ power measurements $P_1, P_2, \ldots, P_n$, the order statistics are $P_{(1)}, P_{(2)}, \ldots, P_{(n)}$, where $P_{(r)}$ is the $r$th largest value.  We write $P_{\max} = P_{(n)}$ to denote the maximum value.  Additionally, for a random variable, $X$, we denote its CDF as $F_X(x)$, its probability density function (PDF) as $f_X(x)$, and the expected value as $\text{E}[X]$.

For a given set of power measurements, the PAPR is defined as the peak power value divided by the mean power, i.e., $\text{PAPR} = P_{\max}/\text{E}[P]$.  Similarly, the CF is defined as $\text{CF} = A_{\max}/\sqrt{\text{E}[P]}$, where $A_{\max} = \sqrt{P_{\max}}$, i.e., $\text{PAPR} = \text{CF}^2$.  Below, it will be convenient to utilize scaled versions of $P$ and $A$ defined as $\widetilde{P} = P/\text{E}[P]$ and $\widetilde{A} = A/\sqrt{\text{E}[P]}$ so that $\text{PAPR} = \widetilde{P}_{\max}$ and $\text{CF} = \widetilde{A}_{\max}$.

We begin with a review of the established distribution theory for PAPR and CF.  To the best of our knowledge, the subsequent theoretical results are new unless stated otherwise.

\subsection{Distribution Theory}
\label{sec:distro_theory}

Since $X_I \sim \mathcal{N}(0,\sigma^2)$ and $X_Q \sim \mathcal{N}(0,\sigma^2)$, it follows that the amplitude, $A$, has a Rayleigh distribution with scale parameter, $\sigma$, and the instantaneous power, $P$, has an exponential distribution with rate parameter, $\lambda = 1/(2\sigma^2)$ \cite{Siddiqui1962, Johnson1994}.  Here, we indicate this with the notation $A \sim \text{Rayleigh}(\sigma)$ and $P \sim \text{Exp}(1/(2\sigma^2))$, where the quantity in parentheses denotes the parameter value.

For $x\geq 0$, the CDFs for $A$ and $P$ are $F_A(x) = 1 - \exp[-x^2/(2\sigma^2)]$ and $F_P(x) = 1 - \exp[-\lambda x] = 1 - \exp[-x/(2\sigma^2)]$, and the mean values are $\text{E}[A] = \sigma \sqrt{\pi/2}$ and $\text{E}[P] = 1/\lambda = 2 \sigma^2$ \cite{Johnson1994}.  Applying the CDF transformation theorem for an increasing function of a random variable \cite[Theorem 2.1.3]{Casella2002}, it follows that for $x\geq 0$, the CDFs of $\widetilde{A} = A/(\sigma \sqrt{2})$ and $\widetilde{P} = P/(2\sigma^2)$ are $F_{\widetilde{A}}(x) = 1 - \exp[-x^2]$ and $F_{\widetilde{P}}(x) = 1 - \exp[-x]$.  In terms of the previously introduced distribution notation, $\widetilde{A} \sim \text{Rayleigh}(1/\sqrt{2})$ and $\widetilde{P} \sim \text{Exp}(1)$, i.e., $\sigma=1/\sqrt{2}$ and $\lambda = 1$ for the Rayleigh and exponential distributions, respectively.    

Suppose we have $n$ i.i.d. measurements of $X_I \sim \mathcal{N}(0,\sigma^2)$ and $X_Q \sim \mathcal{N}(0,\sigma^2)$, yielding $n$ i.i.d. measurements, $\widetilde{P}_1, \widetilde{P}_2, \ldots, \widetilde{P}_n$, of $\widetilde{P} \sim \text{Exp}(1)$ and $n$ i.i.d. measurements, $\widetilde{A}_1, \widetilde{A}_2, \ldots, \widetilde{A}_n$, of $\widetilde{A}\sim \text{Rayleigh}(1/\sqrt{2})$. 
By a well-known formula for the CDF of the maximum of $n$ i.i.d. samples from a known CDF \cite[Eq. 2.1.1]{David1981}, the CDFs of PAPR and CF are
\begin{equation}
    F_{\text{PAPR}}(x) = (1-\exp[-x])^n
\label{eq:PAPR_CDF}
\end{equation}
and
\begin{equation}
    F_{\text{CF}}(x) = (1-\exp[-x^2])^n
\label{eq:CF_CDF}
\end{equation}
for $x\geq 0$. Note that the distributions of PAPR and CF do not depend on $\sigma$, i.e., they are independent of the mean power.  The CDFs (\ref{eq:PAPR_CDF}) and (\ref{eq:CF_CDF}) were previously given in \cite{Ochiai2001}.  

From the CDFS it follows that the quantile functions (inverse CDFs) of PAPR and CF are
\begin{equation}
    Q_{\text{PAPR}}(p) = -\ln(1-p^{1/n})
\end{equation}
and
\begin{equation}
    Q_{\text{CF}}(p) = \sqrt{-\ln(1-p^{1/n})}
\label{eq:CF_quantile_function}
\end{equation}
for $ 0 \leq p < 1$.  The above expressions are useful for engineering design when it is desired to know specific percentiles of the distributions.

Also, differentiating the CDFs yields corresponding expressions for the respective PDFs.  Namely, the PDF of PAPR is
\begin{equation}
    f_{\text{PAPR}}(x) = n\exp[-x](1-\exp[-x])^{n-1}
\label{eq:PAPR_PDF}
\end{equation}
To express the PAPR PDF on a decibel (dB) scale, we apply the PDF transformation theorem \cite[Theorem 2.1.5]{Casella2002} with $\text{PAPR}_{\text{dB}} = 10\log_{10}(\text{PAPR})$ to obtain
\begin{equation}
    f_{\text{PAPR$_{\text{dB}}$}}(y) = \frac{n}{10}\ln(10)
    10^{\frac{y}{10}}e^{-10^{y/10}}
    \left(1-e^{-10^{y/10}}\right)^{n-1}
\label{eq:PAPR_PDF_dB}
\end{equation}

\subsection{Asymptotic Distributions}
\label{sec:asymtptoic_distributions}

From extreme value theory \cite{Castillo1988}, it is known that the distribution for the maximum of an i.i.d. sample from an exponential or Rayleigh distribution asymptotically approaches a Gumbel distribution \cite[Sec.~5.2]{Castillo1988} as the sample size increases.  Specifically, letting $G(z) = \exp[-\exp(-z)]$ denote the standard Gumbel CDF, $\lim_{n\rightarrow \infty} F_{\text{PAPR}}(\ln(n) + z) = G(z)$ and $\lim_{n\rightarrow \infty} F_{\text{CF}}(\sqrt{\ln n} + z/(2\sqrt{\ln n})) = G(z)$ \cite[p.~105-108]{Castillo1988}.  Hence, for large $n$
\begin{equation}
    F_{\text{PAPR}}(x) \approx \exp[-\exp(-(x - \ln n))]
    \label{eq:PAPR_gumbel}
\end{equation}
and 
\begin{equation}
    F_{\text{CF}}(x) \approx \exp[-\exp(-(x - \sqrt{\ln n})2\sqrt{\ln n})].
    \label{eq:CF_gumbel}
\end{equation}
The above CDFs take the form of a general Gumbel distribution, which has CDF $F(x; a, b) = \exp[-\exp(-(x-a)/b)]$ and mean $a + \gamma b$, where $\gamma =  0.5772156649\ldots$ is Euler's constant \cite[p.~185]{Castillo1988}.

\subsection{Mean Peak-to-Average Power Ratio}
\label{sec:mean_PAPR}

First, we obtain an approximate asymptotic expression for the mean PAPR from (\ref{eq:PAPR_gumbel}) and the formula for the mean of the Gumbel distribution.  Namely, for large $n$ 
\begin{equation}
    \text{E}[\text{PAPR}] \approx \ln n + \gamma.
\label{eq:mean_PAPR_gumbel_approx}
\end{equation}

An exact closed-form formula for the mean PAPR is derived from established results on order statistics for the standard exponential distribution, which imply that the mean of the $r$th order statistic, $\widetilde{P}_{(r)}$, is \cite[Eq.~19.12]{Johnson1994} 
\begin{equation}
    \text{E}[\widetilde{P}_{(r)}]  = \sum_{j=1}^r \frac{1}{n-j+1} = \sum_{k=n-r+1}^{n}\frac{1}{k}.
\end{equation}
Setting $r=n$ for the maximum value, $\widetilde{P}_{\max} = \text{PAPR}$, gives
\begin{equation}
    \text{E}[\text{PAPR}]  =  H_n
\label{eq:mean_PAPR}
\end{equation} 
where $H_n = \sum_{k=1}^{n} 1/k$ denotes the $n$th harmonic number.  The above formula is our main result and will be applied later in Sections~\ref{sec:assumption_violations} and \ref{sec:spectrogram}.     

Efficient computation of $H_n$ for large $n$ can be carried out with the asymptotic approximation \cite[p.~480]{Graham1994}
\begin{equation}
    H_n \approx \ln{n} + \gamma + \frac{1}{2n} - \frac{1}{12n^2} + \frac{1}{120 n^4}.
\label{eq:asymptotic_form}
\end{equation}
The first two terms, which correspond to (\ref{eq:mean_PAPR_gumbel_approx}), are sufficient to ensure that the relative approximation error is less than $0.1\%$ when $n\geq 100$.

\subsection{Mean Crest Factor}
\label{sec:mean_CF}
First, from (\ref{eq:CF_gumbel}) and the formula for the mean of a Gumbel distribution, it follows that for large $n$, the mean CF is roughly 
\begin{equation}
    \text{E}[\text{CF}] \approx \sqrt{\ln n} + \gamma/(2\sqrt{\ln n}).
\label{eq:mean_CF_gumbel_approx}
\end{equation}
The above approximation was found in \cite[Eq. (59)]{Longuet1952} by other means in the context of narrow-band sea wave maxima.  

We can also use (\ref{eq:mean_PAPR}) to derive an upper bound for the mean CF.  Since $\text{CF} = \sqrt{\text{PAPR}}$, and because square root is a concave function, Jensen's inequality \cite[Thm.~4.7.7]{Casella2002} implies that $\text{E}[\text{CF}] \leq \sqrt{\text{E}[\text{PAPR}]}$.  Thus, (\ref{eq:mean_PAPR}) yields 
\begin{equation}
\text{E}[\text{CF}] \leq \sqrt{H_n}.  
\label{eq:CF_bound}
\end{equation} 

An exact closed-form expression for the mean CF can be obtained from known results for the mean order statistics of a Rayleigh distributed sample.  Namely, starting with \cite[Eq. 18.82]{Johnson1994} and simplifying yields 
\begin{equation}
\text{E}[\text{CF}] = \frac{\sqrt{\pi}}{2}\sum_{k=1}^n \binom{n}{k} (-1)^{k-1} k^{-1/2}
\label{eq:CF_exact1}
\end{equation}
where $\binom{n}{k}$ denotes the binomial coefficient.  Unfortunately, the above formula is not useful for computation when $n > 50$ (approximately), due to numerical error.  

Instead, we express the mean CF as an integral that can be accurately evaluated for large $n$ with standard numerical quadrature routines.  The probability integral transformation \cite[Th. 2.1.10]{Casella2002} implies that the random variable obtained by transforming CF with its CDF, $U = F_{\text{CF}}(\text{CF})$, follows a standard uniform distribution on the interval $[0,1]$. Thus, transforming a standard uniform random variable, $U$, with the CF quantile function (inverse CDF) yields $Q_{\text{CF}}(U) = \text{CF}$, and hence, $\text{E}[\text{CF}] = \text{E}[Q_{\text{CF}}(U)]$.  Writing the last expectation in integral form with (\ref{eq:CF_quantile_function}) gives
\begin{equation}
\text{E}[\text{CF}] = \int_0^1 \sqrt{-\ln(1-u^{1/n})}\,du.
\label{eq:CF_exact2}
\end{equation}

\begin{figure}[tb]
    \centering
    \subfloat[\label{subfig:rel_err_CF}]{
        \includegraphics[width = \columnwidth]{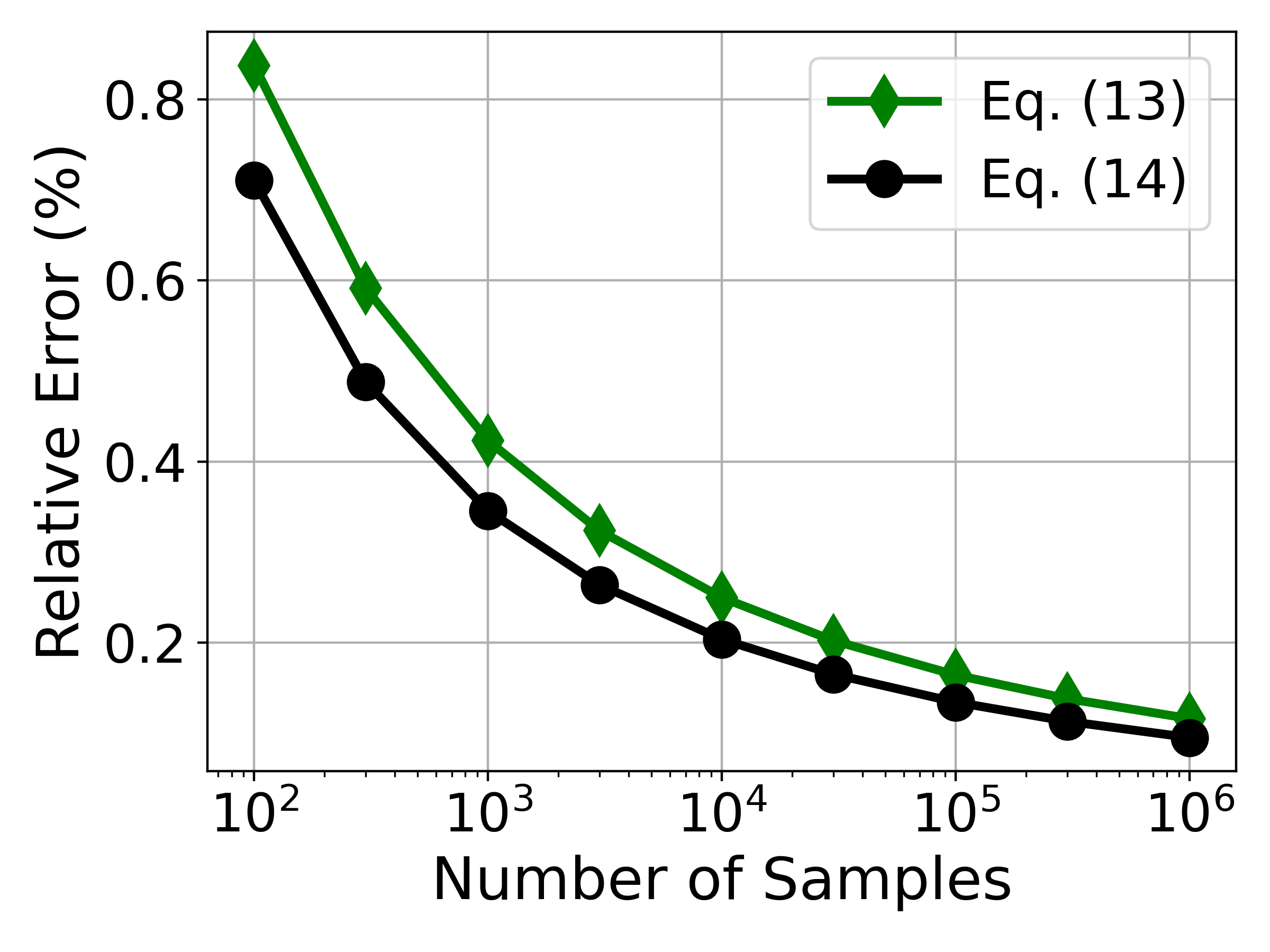}}
    \par
    \medskip
    \subfloat[\label{subfig:rel_err_PAPR}]{
        \includegraphics[width = \columnwidth]{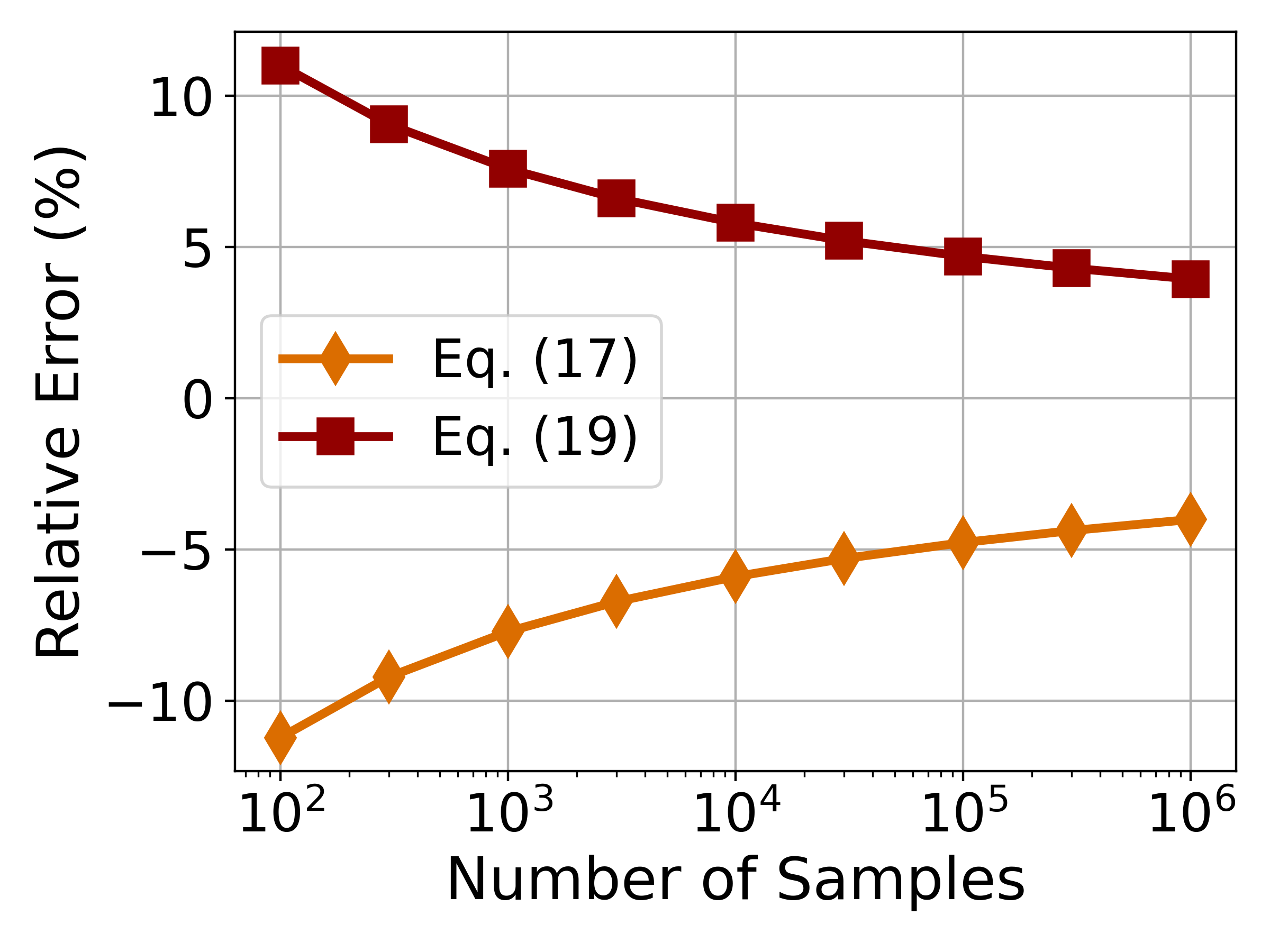}}
    \caption{Relative error in approximations for mean CF and mean PAPR with WGN I/Q data. (a) Mean CF approximations (see Section~\ref{sec:mean_CF}). (b) Mean PAPR approximations (see Section~III). }
    \label{fig:rel_error_plots}
\end{figure}

Fig.~\ref{subfig:rel_err_CF} shows the relative error of the approximation (\ref{eq:mean_CF_gumbel_approx}) and the bound (\ref{eq:CF_bound}) (the lower plot, Fig.~\ref{subfig:rel_err_PAPR}, is discussed in the next section).  The relative error, in percent, is calculated as $\text{Rel Err} = (\text{approx} - \text{exact})/\text{exact} \times 100$, where $\text{approx}$ denotes the values from (\ref{eq:mean_CF_gumbel_approx}) and (\ref{eq:CF_bound}) and $\text{exact}$ denotes the value from (\ref{eq:CF_exact2}), all expressed on a linear scale.  The definite integral in (\ref{eq:CF_exact2}) was evaluated using the `integrate.quad' method in the SciPy python library \cite{SciPy2020}\footnote{Certain equipment, instruments, software, or materials are identified in this paper in order to specify the experimental procedure adequately.  Such identification is not intended to imply recommendation or endorsement of any product or service by NIST, nor is it intended to imply that the materials or equipment identified are necessarily the best available for the purpose.} with the relative error tolerance parameter set to $10^{-10}$.  From this plot, we see that (\ref{eq:mean_CF_gumbel_approx}) and (\ref{eq:CF_bound}) are excellent approximations for the mean CF, with (\ref{eq:CF_bound}) being slightly better.  Thus, (\ref{eq:mean_CF_gumbel_approx}) and (\ref{eq:CF_bound}) are a simple, accurate alternatives to numerically evaluating the integral in (\ref{eq:CF_exact2}).  

\section{Comparison with Previously Published Approximate Formulas for Mean PAPR}
\label{sec:approx_mean_PAPR}

We compare the exact result (\ref{eq:mean_PAPR}) presented in Section~\ref{sec:mean_PAPR} to two approximate formulas from the engineering literature for the mean PAPR of WGN in I/Q data.  The first formula, given in a book on microwave measurements \cite[p.~588]{Dunsmore2020}, states that for noise-like signals, the expected PAPR, expressed in linear units, is roughly 
\begin{equation}
    \text{E}[\text{PAPR}] \approx \ln(n)
    \label{eq:DunsmorePAPR}
\end{equation}
where $n$ is the number of samples.  This formula is equivalent to the first term of the asymptotic expansion (\ref{eq:asymptotic_form}).  Thus, it underestimates the expected PAPR, with the difference approaching $\gamma \approx 0.5772$ as $n$ approaches infinity.  

An equipment vendor application note gives the following formula, expressed in linear units, for the average PAPR found from ``a combination of analysis, approximation and experimentation'' \cite[p.~17-18]{Keysight2021}
\begin{equation}
    \text{E}[\text{PAPR}] \approx \ln(2\pi \tau \text{BW}_i + e)
\label{eq:Keysight1}
\end{equation}
where $e = 2.71828 \ldots$ is Euler's number, $\tau$ is the observation period and $\text{BW}_i$ is the one-sided impulse bandwidth for the resolution bandwidth (RBW) filter.  Ref. \cite[p.~7]{Cotton2014} applies the above formula in the context of spectrum monitoring.  

To express Eq.~(\ref{eq:Keysight1}) in terms of the number of samples, we assume that the one-sided impulse bandwidth is well approximated by one half the sampling rate, $f_s$, i.e., the signal is critically sampled.\footnote{In practice, an anti-aliasing filter may reduce the bandwidth further.}  Multiplying by the observation period, then yields  $\tau \text{BW}_i \approx \tau f_s/2 = n/2$, where $n$ is the number of samples. Hence, we obtain 
\begin{equation}
\text{E}[\text{PAPR}] \approx \ln(\pi n + e).
\label{eq:Keysight2}
\end{equation}  

In order to compare the above formula to (\ref{eq:mean_PAPR}), observe that since $\pi n + e = \pi n(1+ e/(\pi n))$, 
\begin{align}
     \ln(\pi n + e) &= \ln(\pi n) + \ln\left(1 + \frac{e}{\pi n}\right) \\
     &\approx \ln n + \ln \pi + \frac{e}{\pi n}
\end{align}
where the last approximation follows from the first-order Maclaurin expansion for $\ln(1+x)$.  Finally, since the last term approaches zero as $n$ goes to infinity, we find that (\ref{eq:Keysight2}) is roughly $\text{E}[\text{PAPR}] \approx \ln n + \ln \pi$ for large $n$.  Thus, compared with the asymptotic expansion (\ref{eq:asymptotic_form}), (\ref{eq:Keysight2}) overestimates the mean PAPR, with the error approaching $\ln \pi - \gamma \approx 0.5675$ as $n$ approaches infinity.  

Fig.~\ref{subfig:rel_err_PAPR} shows the relative difference, in percent, between the approximate expressions (\ref{eq:DunsmorePAPR}) and (\ref{eq:Keysight2}) and the exact result (\ref{eq:mean_PAPR}). These plots indicate that the approximate formulas result in relative errors ranging from greater than $10\%$ for $n=100$ to less than $5\%$ for $n>10^5$.  As expected, (\ref{eq:DunsmorePAPR}) and (\ref{eq:Keysight2}) underestimate and overestimate the true value, respectively.  Even on a decibel scale, these errors can be significant for some applications.  For example, for $n=1000$, the approximate formulas yield errors of roughly $\pm 0.3$\,dB.

\section{Examples of WGN Assumption Violations}
\label{sec:assumption_violations}

Our theoretical findings are built upon assumptions about the nature of the noise and the operational space of the instruments recording it.  Namely, our results require that the recorded samples of I and Q components are zero-mean WGN, implying that the I and Q components satisfy the following conditions:
\begin{itemize}
    \item[(1)] not dominated by nonlinear distortion effects, such as quantization noise 
    \item[(2] balanced, i.e., the gains of each RF leg are equal, and the phase difference is $90^{\circ}$
    \item[(3)] statistically independent, white Gaussian processes.
\end{itemize}

Below, we provide examples illustrating how breakdowns in the above assumptions can occur in practical measurements of thermal noise and impact the expected PAPR.  Conversely, when measuring thermal noise, significant deviations from the expected mean PAPR for WGN provide evidence that I/Q measurements are non-ideal and violate one or more of the above assumptions.     

\subsection{Nonlinear Distortion}
\label{subsec:nonlinear_distortion}

To illustrate the impact of nonlinear distortions arising from improper selection of reference level (the maximum expected input power), we collected repeated RF measurements with a National Instruments PXIe-5646 vector signal transceiver (VST), which has a 14-bit analog-to-digital converter (ADC) resolution \cite{VSTspecs}.  The RF input to the VST was terminated with a matched, $50\,\Omega$ load so that the input signal consisted of room-temperature thermal noise.  Repeated captures of I/Q samples were collected with a center frequency of $5$\,GHz at a sampling rate of $30$~Megasamples per second (MS/s) with four different VST reference level settings: $-50$\,dBm, $10$\,dBm, $20$\,dBm, and $30$\,dBm. The $-50$\,dBm reference level was sufficiently low so as to avoid quantization effects, whereas the other reference levels were sufficiently large to ensure impacts from nonlinear quantization. 

\begin{figure}[t]
    \centering
    \includegraphics[width = \columnwidth]{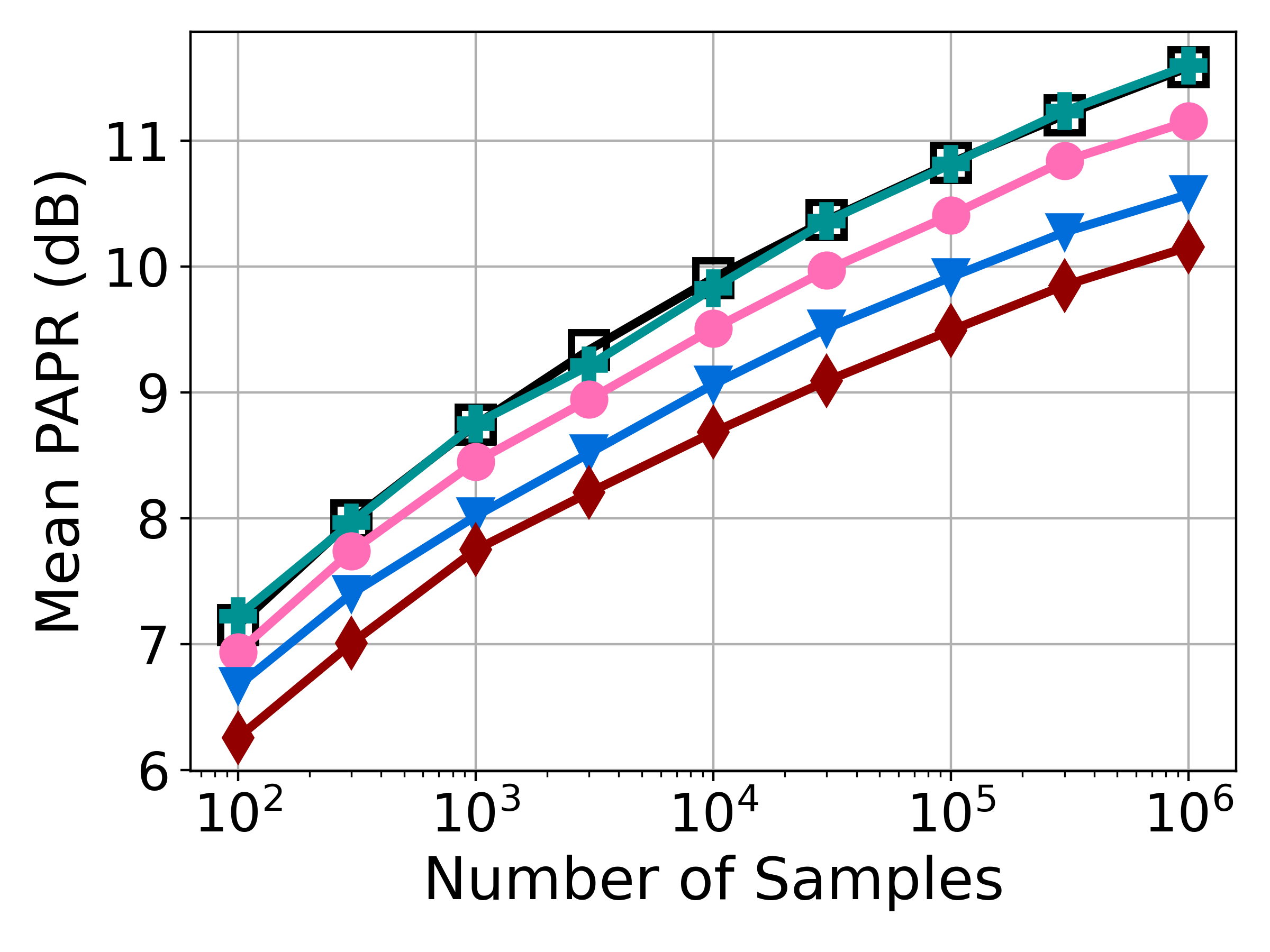}
    \includegraphics[width = \columnwidth]{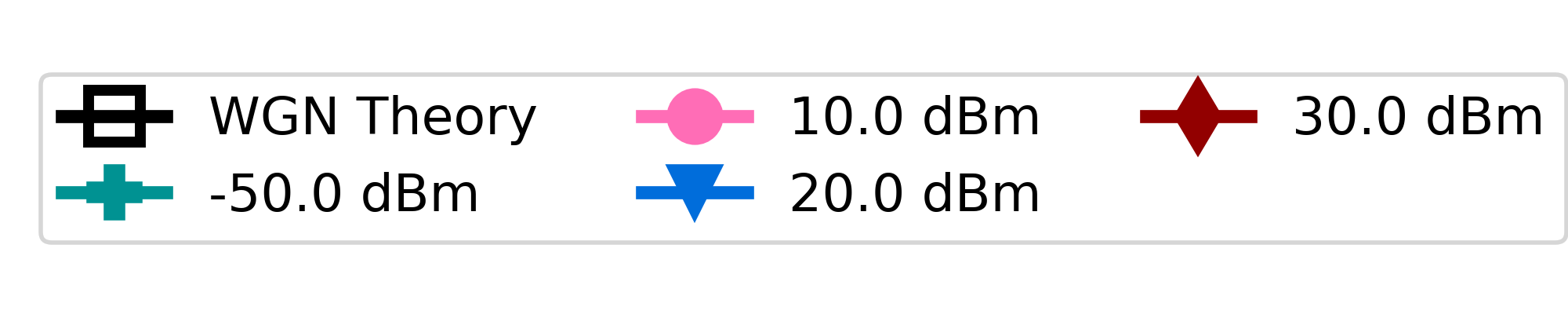}
    \includegraphics[width = \columnwidth]{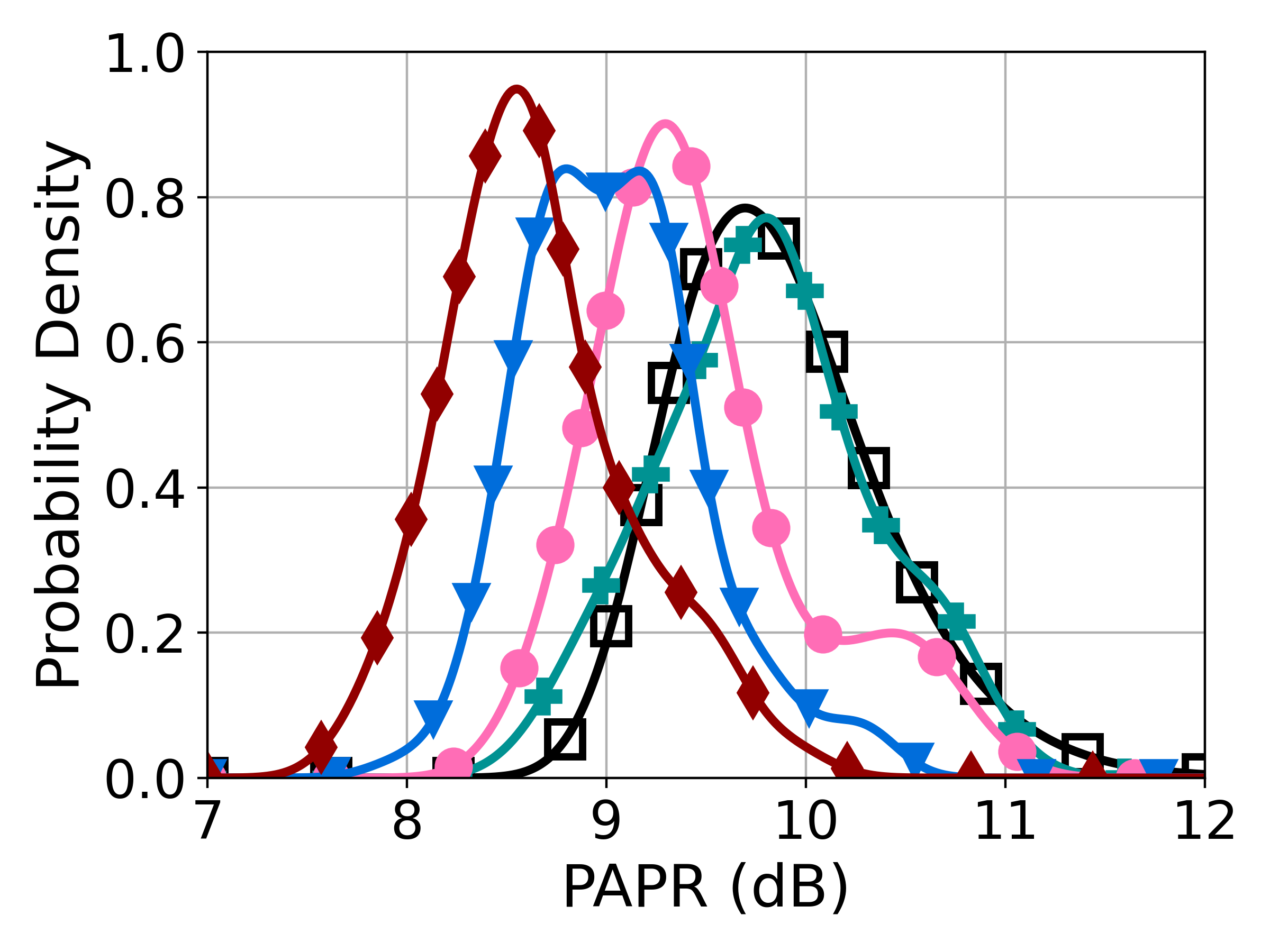}
    \caption{Vector signal transceiver (VST) measurements comparing the PAPR of I/Q thermal noise measurements with various reference levels.  Top: Mean PAPR vs sample size. Bottom: Estimated PDFs for $n=10^4$ samples.}
    \label{fig:PAPR_VST}
\end{figure}

Fig.~\ref{fig:PAPR_VST}~(top) plots the mean PAPR estimated from 120 repeated measurements at each sample size for each reference level setting, as well as the theoretical value for I/Q WGN predicted by (\ref{eq:mean_PAPR}).  Error bars are suppressed, since in all cases $95$\% confidence intervals were smaller than $\pm 0.19$\,dB.  From this plot, we see that when the VST reference level was set to -50\,dBm, the empirical mean PAPR was very close to the theoretical mean value for WGN.  By contrast, when the VST reference level was set to the higher levels, the recorded noise was influenced by nonlinear distortion stemming from quantization effects and the resulting mean PAPR was lower than for WGN.  Fig.~\ref{fig:PAPR_VST}~(bottom) plots empirical PDFs of PAPR for the $n=10^4$ case, estimated with the nonparametric kernel density estimation (KDE) method and compares them to the theoretical PDF given by (\ref{eq:PAPR_PDF_dB}).  KDE was implemented using `stats.gaussian\_kde' in the SciPy Python library \cite{SciPy2020} with default settings, which apply Gaussian kernels with the bandwidth parameter set with Scott's rule \cite{Scott1992}.  As expected, the empirical PDF for the $-50$\,dBm reference level had excellent agreement with the theoretical PDF (\ref{eq:PAPR_PDF_dB}).

\subsection{I/Q Imbalance}

In RF transmitters and receivers, I/Q imbalance (or mismatch) results from gain and phase mismatches on the I and Q branches \cite[Sec.~2.7]{Smaini2012}.  Let $\Delta \phi$ be the phase mismatch between the I and Q components, and let $\Delta_G = |G_I-G_Q|/G_Q$ be the relative gain mismatch, where $G_I$ and $G_Q$ are the gains for the I and Q branches.  Also, let $A = 1 - \Delta G/2$ and $B = 1 + \Delta G/2$.  Using matrix multiplication, I/Q imbalance in an RF receiver can be modeled as \cite[Eq.~(2.131)]{Smaini2012}
\begin{equation}
    \begin{pmatrix}
        Y_I \\
        Y_Q
    \end{pmatrix} = 
    \begin{pmatrix}
        A\cos(\Delta \phi/2) &A\sin(\Delta \phi/2) \\
        B\sin(\Delta \phi/2) &B\cos(\Delta \phi/2) 
    \end{pmatrix}
    \begin{pmatrix}
        X_I \\
        X_Q
    \end{pmatrix}
\label{eq:IQ_imbalance_model}
\end{equation}
where $X_I$ and $X_Q$ are the I/Q components of the ideal (balanced) baseband signal, and $Y_I$ and $Y_Q$ are the I/Q components of the baseband signal with I/Q imbalance.  

\begin{figure}[t]
    \centering
    \includegraphics[width = \columnwidth]{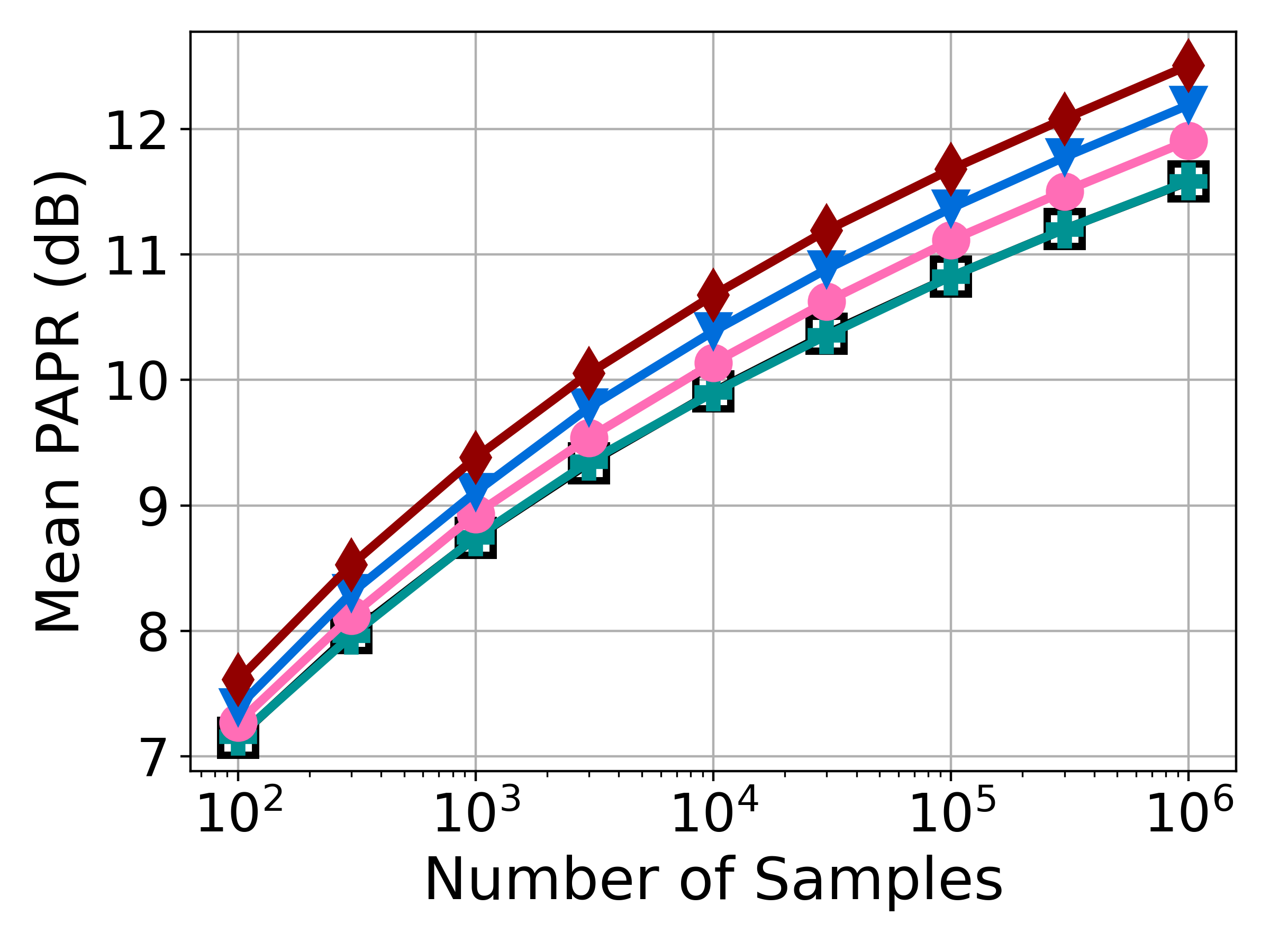}
      \includegraphics[width = \columnwidth]{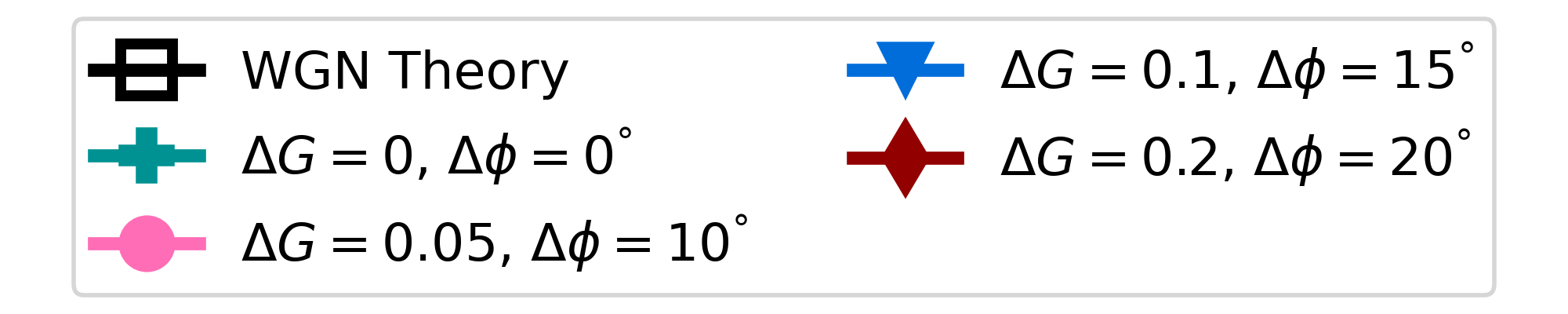}
    \includegraphics[width = \columnwidth]{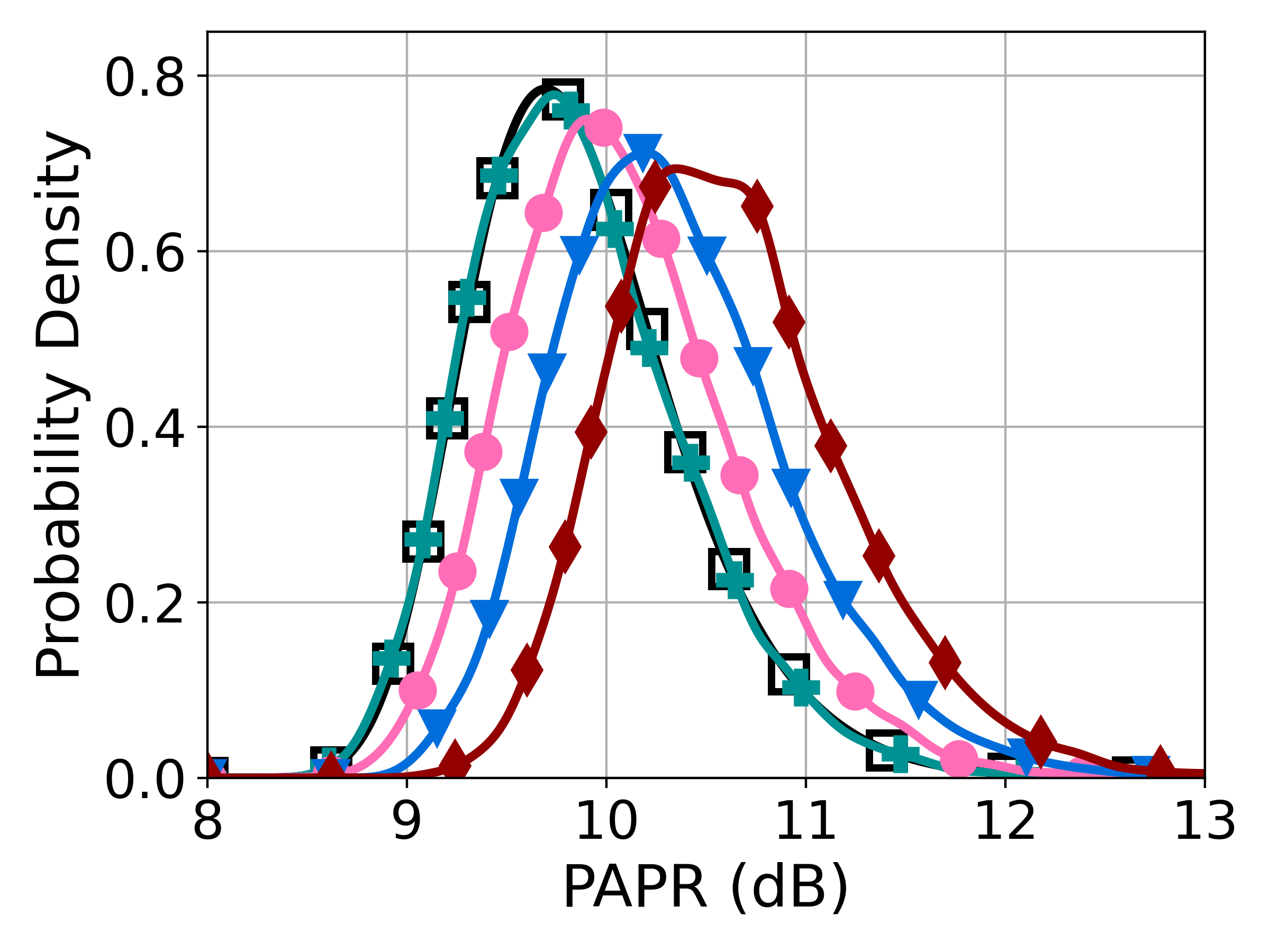}
    \caption{Simulated I/Q Imbalance Example.  Top: Mean PAPR vs. sample size. Bottom: Estimated PDFs for $n=10^4$ samples.}
    \label{fig:PAPR_IQ_imbalance}
\end{figure}

Monte Carlo simulations were performed to demonstrate the impacts of I/Q imbalance on the expected PAPR.  For each sample size, I/Q samples were independently drawn from standard normal $\mathcal{N}(0,1)$ distributions and transformed with (\ref{eq:IQ_imbalance_model}) for $(\Delta G, \Delta \phi) = (0.05, 10^\circ), (0.1, 15^\circ)$, and $(0.2, 20^\circ)$.  In addition, the case of no I/Q imbalance, $(\Delta G, \Delta \phi) = (0, 0)$, was simulated for baseline comparison.  Mean PAPR estimated from $10^4$ independent Monte Carlo trials for each sample size is plotted in Fig.~\ref{fig:PAPR_IQ_imbalance}~(top) together with the theoretical value (\ref{eq:mean_PAPR}) for ideal I/Q WGN.  In all cases, the 95\% confidence intervals were smaller than $\pm 0.022$\,dB.  From this plot, we observe that the expected PAPR increased as the I/Q imbalance became more severe.  For example, in the $(\Delta G, \Delta \phi) = (0.2, 20^\circ)$ case with $n=10^4$ samples, the mean PAPR increased by roughly $0.8$\,dB over the baseline case with no I/Q imbalance.  Fig.~\ref{fig:PAPR_IQ_imbalance}~(bottom) plots PDFs for the $n=10^4$ case estimated with the KDE method described in Section~\ref{subsec:nonlinear_distortion} and compares them with the theoretical PDF for ideal I/Q WGN given by (\ref{eq:PAPR_PDF_dB}).  For both plots in Fig.~\ref{fig:PAPR_IQ_imbalance} in the ideal case with no I/Q imbalance, there was excellent agreement between the Monte Carlo results and theory.   

\subsection{Correlated Samples}
\label{sec:LPFWN}

Finally, we examined the impacts of I/Q measurement correlations on the expected PAPR. Specifically, Monte Carlo simulations were performed to assess PAPR of correlated samples generated by low-pass filtering I/Q WGN.  This example gives insight into the case of band-limited Gaussian noise when the (one-sided) bandwidth is smaller than the Nyquist frequency, equal to one half the sampling rate.  

\begin{figure}[tb]
    \centering
    \includegraphics[width = \columnwidth]{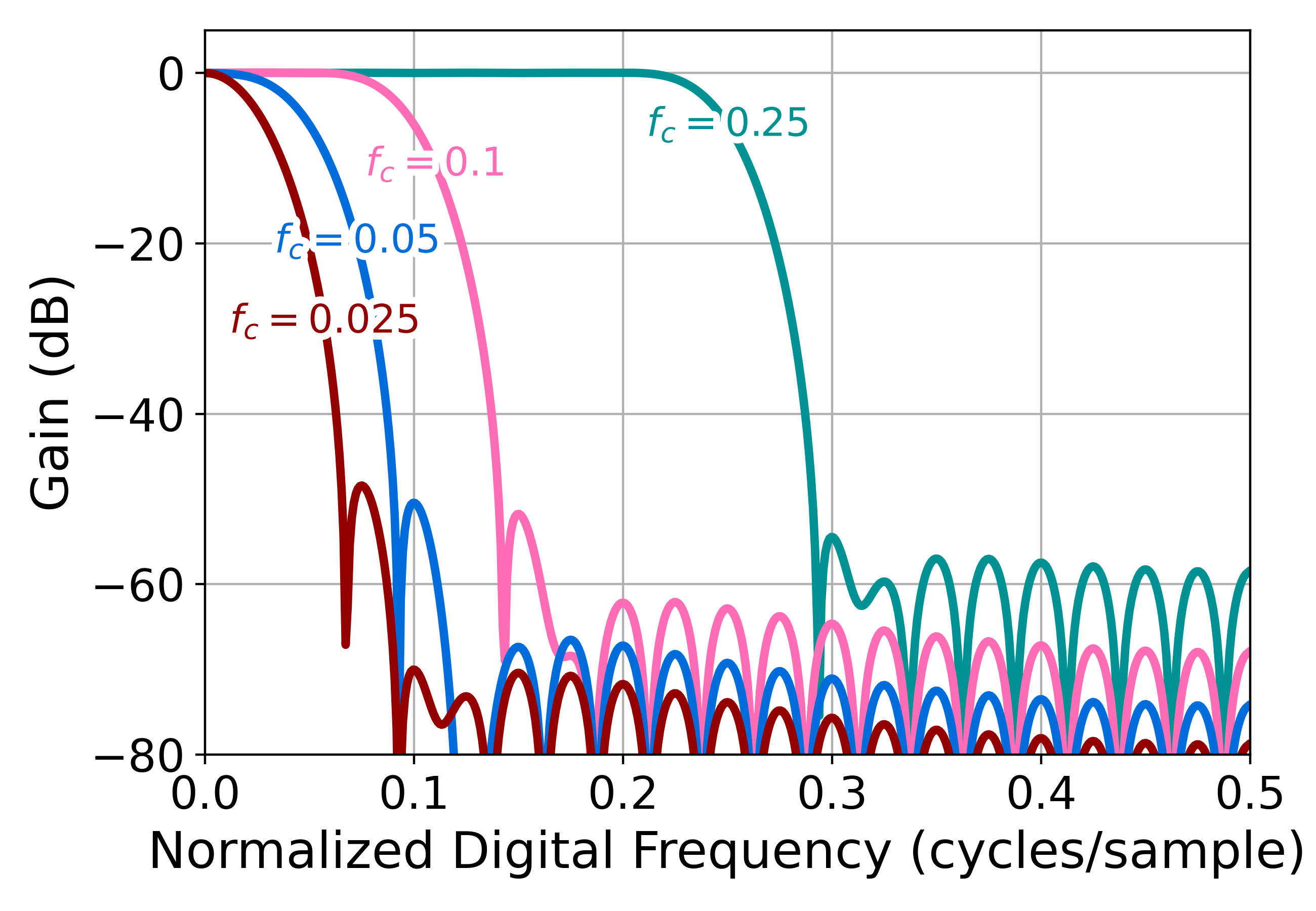}
\caption{Frequency responses of low-pass digital filters used for correlated noise simulations.  The cutoff frequency, $f_c$, is specified in cycles per sample.}
\label{fig:LP_filter_responses}
\end{figure}

For each sample size, I/Q samples were independently drawn from standard normal $\mathcal{N}(0,1)$ distributions and then filtered with a 40th-order (41 tap) low-pass digital finite impulse response (FIR) filter.  The digital FIR filter was designed using the window method with a Hamming window \cite{Mitra2006}, as implemented by the `signal.firwin' method in the SciPy python library \cite{SciPy2020} with default parameters.  To avoid transients at the beginning or end of the length $n$ filtered noise sequence, an additional number of WGN I/Q samples equal to the filter order were added to both ends of the sequence and then discarded after filtering.  The filter cutoff frequency, $f_c$, specified in normalized digital frequency\footnote{In terms of analog frequency, $f$, and sampling rate, $f_s$, the normalized digital frequency is $f/f_s$ and lies in the interval $[0, 0.5]$.} units of cycles per sample, took the values $\{0.025, 0.05, 0.1, 0.25\}$.  Frequency responses for the lowpass FIR filters are shown in Fig.~\ref{fig:LP_filter_responses}.

\begin{figure}[t]
    \centering
    \includegraphics[width = \columnwidth]{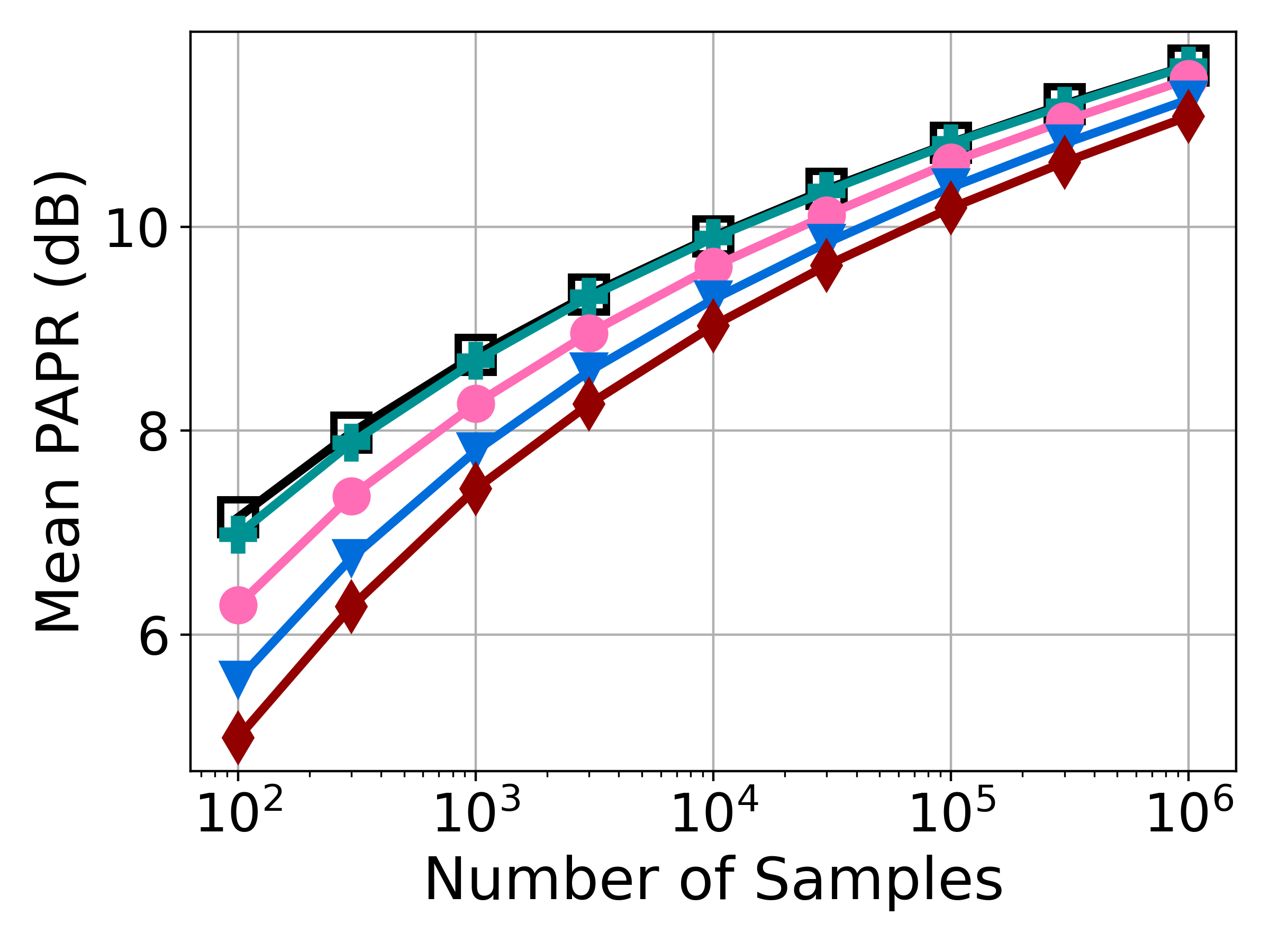}
      \includegraphics[width = \columnwidth]{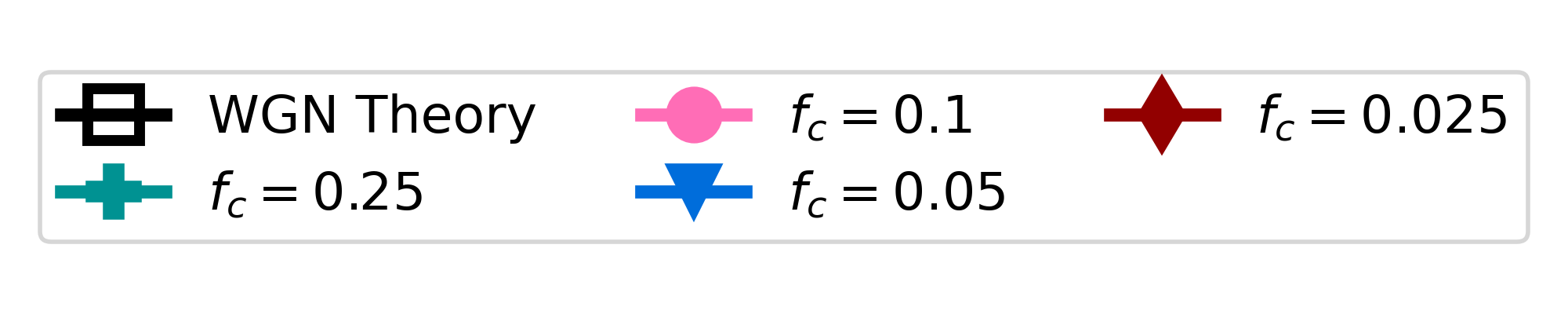}
    \includegraphics[width = \columnwidth]{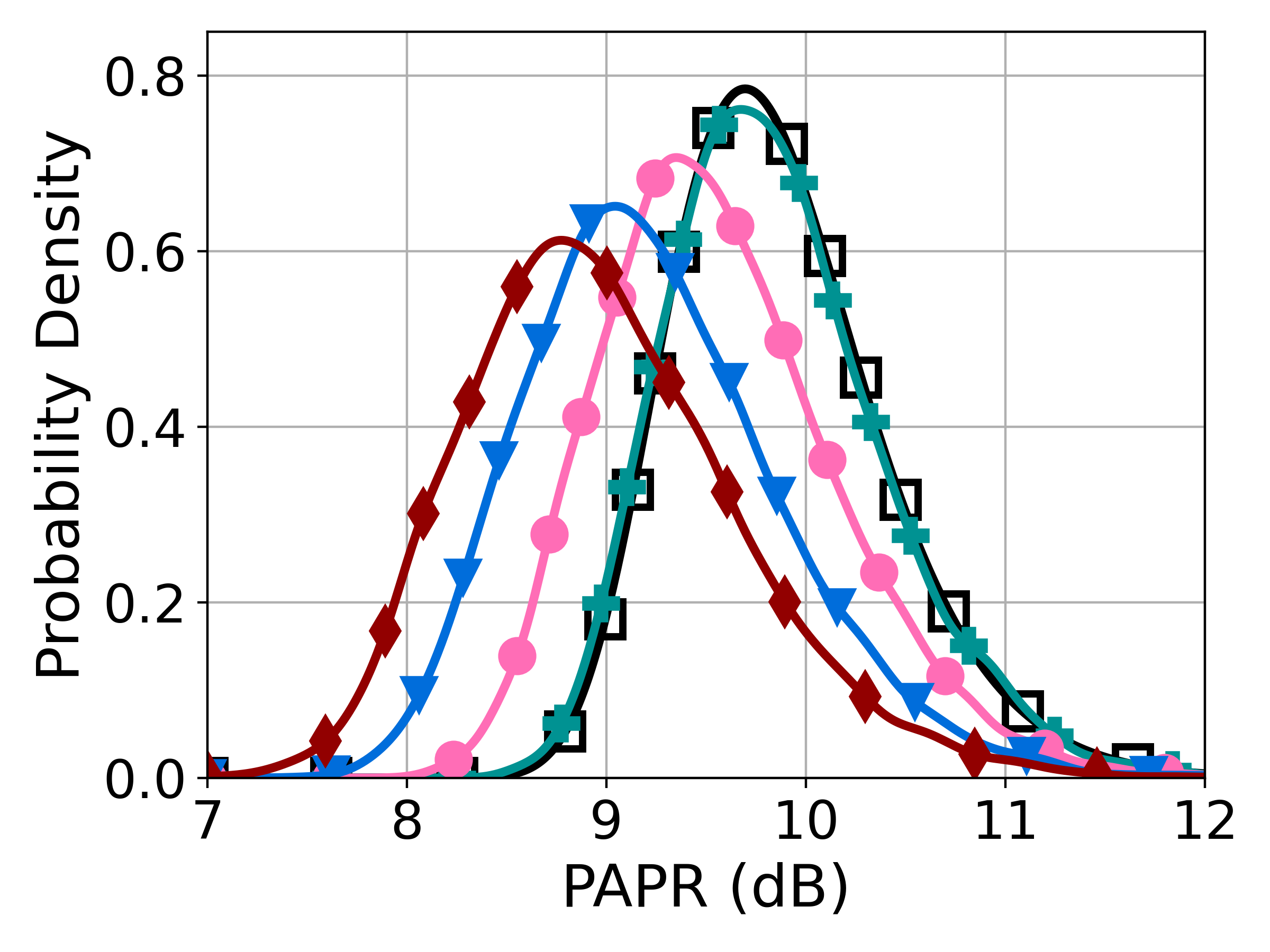}
    \caption{Correlated noise simulations with low-pass filter cutoff frequency, $f_c$, specified in cycles per sample.  (Top) Mean PAPR versus sample size.  (Bottom) Estimated PDFs for $n = 10^4$ samples.}
    \label{fig:PAPR_LPFWN}
\end{figure}

Mean PAPR estimated from $10^4$ independent Monte Carlo trials for each sample size is plotted in Fig.~\ref{fig:PAPR_LPFWN}~(top) together with the theoretical mean PAPR (\ref{eq:mean_PAPR}) for ideal I/Q WGN.  In all cases, 95\% confidence intervals were smaller than $\pm 0.025$\,dB.  We see that the mean PAPR generally decreased as the filter cutoff, $f_c$, decreased.  Moreover, the mean PAPR was very close to the theoretical WGN value for $f_c \geq 0.25$, indicating that (\ref{eq:mean_PAPR}) is an accurate approximation when $0.25 \leq f_c \leq 0.5$.  Fig.~\ref{fig:PAPR_LPFWN}~(bottom) compares PDFs estimated with the KDE method described in Section~\ref{subsec:nonlinear_distortion} for the $n=10^4$ sample size to the I/Q WGN theoretical PDF (\ref{eq:PAPR_PDF_dB}).

\section{Example Application to Spectral Analysis}
\label{sec:spectrogram}

Finally, we demonstrate how our main result, (\ref{eq:mean_PAPR}), can be applied to spectral analysis of sampled I/Q signals to determine if a given frequency band contains only WGN.  Specifically, we illustrate the application to spectrograms, which are widely used to assess the frequency content of a nonstationary signal as it varies in time.  A spectrogram is the magnitude squared of a sampled short-time Fourier transform (STFT), computed by applying the discrete Fourier transform (DFT) successively to shifted, windowed signal segments \cite[Sec.~15.3]{Mitra2006}.  

The application relies on the fact that the real and imaginary components of the DFT of a windowed realization of a stationary, circularly symmetric complex WGN process are i.i.d. zero-mean WGN \cite{Richards2013}.  In particular, suppose that an N-point discrete-time random process, $y[t]$, can be decomposed as the sum of (a) a random process only supported on a proper subset, $B$, of the discrete-frequency DFT domain and (b) a stationary, circularly symmetric complex WGN process.  Using the linearity of the DFT, it then follows that the real and imaginary components of the DFT of $y[t]$ are i.i.d. zero-mean WGN on the complement of $B$.  Hence, the mean PAPR of the spectrogram of $y[t]$ for each frequency bin not in $B$ is given by (\ref{eq:mean_PAPR}) with $n$ equal to the number of spectrogram time bins.  

To illustrate how the above theoretical observation can be applied, we used a Deepwave Digital AIR-T~7201B \cite{deepwave} software defined radio (SDR) to record I/Q data at a sample rate of 30\,Ms/s, for 30\,ms at a center frequency of 3700~MHz.  A general-purpose antenna, band-pass filter, and low-noise amplifier were used as an RF front-end prior to the SDR to amplify signals in the 3\,GHz to 4\,GHz frequency range.  The amplitude measured by the SDR was not calibrated.  The acquisition band, which spanned 3685\,MHz to 3715\,MHz, covers the transition between the 3550\,MHz to 3700\,MHz Citizens Broadband Radio Service (CBRS) band \cite{CBRS} and the 3700\,MHz to 3980\,MHz band \cite{C-band}, which are both used for mobile wireless communications in the United States. 

\begin{figure}[t]
    \centering
      \includegraphics[width = \columnwidth]{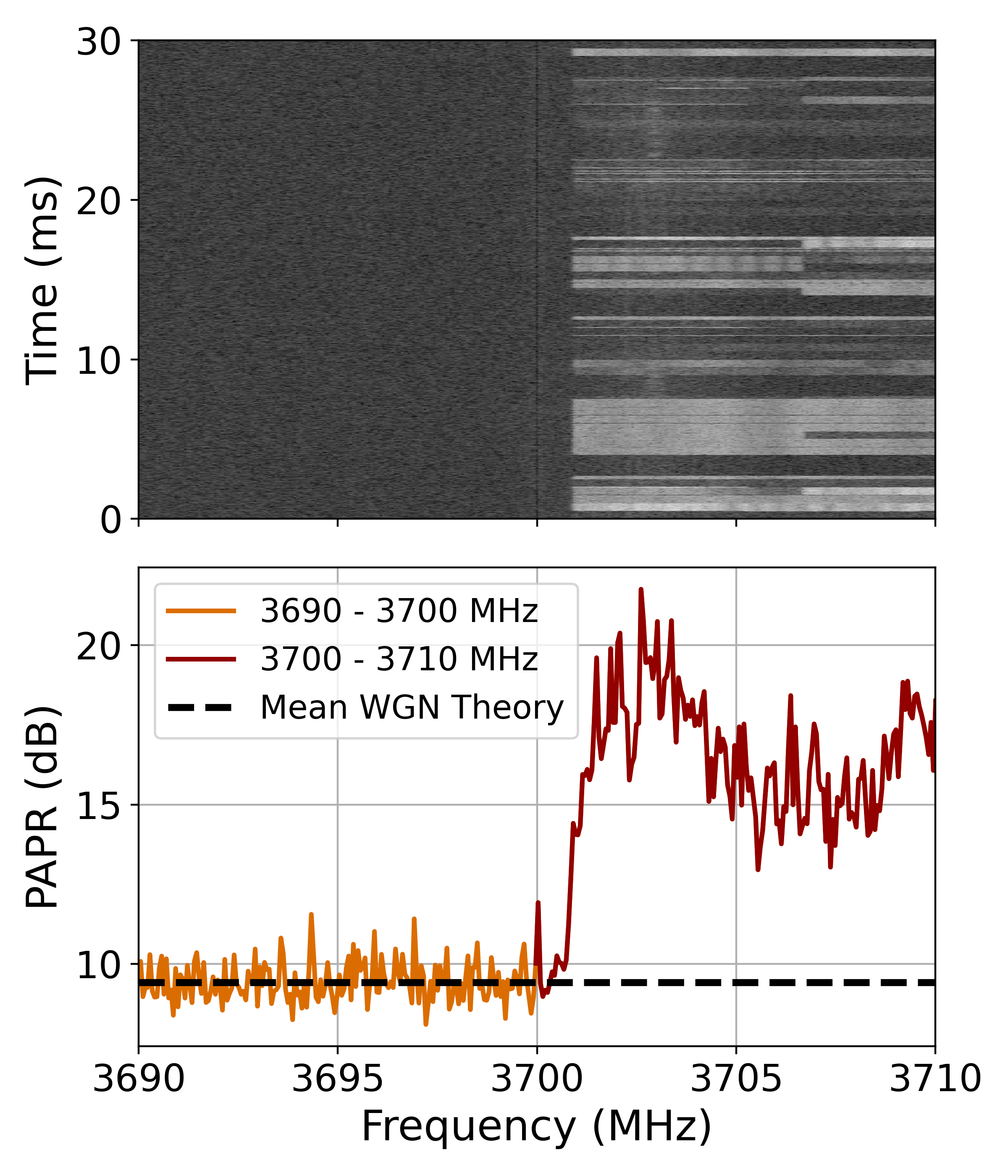}
    \caption{Spectrogram example for an I/Q capture centered at 3.7\,GHz.  Top: Spectrogram with grayscale spanning $-110$\,dB to $-60$\,dB (uncalibrated power units).  Bottom: PAPR vs. frequency.}
    \label{fig:spectrogram}
\end{figure}

A spectrogram computed from the I/Q capture is shown in Fig.~\ref{fig:spectrogram}~(top), with the frequency range cropped to 3690\,MHz to 3710\,MHz for this example.  The spectrogram was calculated with the `signal.ShortTimeFFT.spectrogram' method in the Python SciPy library \cite{SciPy2020}, using a Hann window of length 512 and 50\% window overlap, yielding a total of 512 frequency bins and 3517 time bins.  To suppress the SDR local oscillator (LO) signal at 3700\,MHz, the `constant' detrending option was applied, which subtracted the mean from each window segment.  The spectrogram, which visualizes the (uncalibrated) signal power across time and frequency shows visible mobile wireless emissions in the 3700\,MHz to 3710\,MHz band, but no obvious emissions in the 3690\,MHz to 3700 band.

PAPR was calculated in each spectrogram frequency bin by finding the ratio of the maximum to the mean across all $3517$ time bins and is plotted in Fig.~\ref{fig:spectrogram}~(bottom) on a decibel scale.  The mean PAPR in the $3690$\,MHz to $3700$\,MHz band was estimated to be $9.48$\,dB (95\% confidence interval $[9.38, 9.58]$\,dB) by averaging PAPR across the 170 frequency bins covering that band.  The mean PAPR given by (\ref{eq:mean_PAPR}) with $n=3517$, was found to be $9.42$\,dB, as indicated by the thick dashed line in Fig.~\ref{fig:spectrogram}~(bottom).  There was strong agreement between the observed mean PAPR in the $3690$\,MHz to $3700$\,MHz band and the theoretical mean value for WGN, supporting the hypothesis that the band contained only thermal noise during the observation interval.  By contrast, the observed PAPR in the $3700$\,MHz to $3710$\,MHz band was significantly higher, consistent with mobile wireless emissions.        

This example illustrates how the theoretical formula (\ref{eq:mean_PAPR}) can be used as a quick, simple method to determine if a given frequency band contains only thermal noise, which is useful for signal detection and channel occupancy evaluations.  We make three observations.  First, the approach is robust to the choice of spectrogram window function.  Namely, the effect of windowing cancels out in the PAPR since in the case of WGN, the value of each spectrogram frequency bin follows an exponential distribution with the distribution parameter scaled by the energy in the window sequence \cite[Eq.~44]{Richards2013}.  Second, we have found this method to be fairly robust to the degree of spectrogram window overlap, e.g., 50\%, 75\%, although extreme degrees of windows overlap ($> 90 \%$) induce too much statistical dependence between window segments, rendering (\ref{eq:mean_PAPR}) inaccurate.  Last, since PAPR is generally independent of power calibration factors, this method does not require calibrated I/Q values.

\section{Discussion \& Conclusions}

After reviewing established distribution theory for PAPR and CF of zero-mean, i.i.d. WGN in sampled I/Q data, we presented approximate and exact theoretical expressions for the mean PAPR and mean CF.  In particular, we derived an exact, closed-form formula for the mean PAPR, $(\ref{eq:mean_PAPR})$, and compared it to approximate formulas given in the instrumentation literature, demonstrating that the previously published formulas are asymptotically biased.  Additionally, we used (\ref{eq:mean_PAPR}) to derive a closed-form upper bound to the mean CF and showed that it is a very accurate approximation to the mean CF.  Next, we examined circumstances where violations of the WGN assumptions may arise in practical I/Q measurements of thermal noise and lead to deviations from the expected PAPR.  Consequently, when measuring thermal noise, significant deviations from the expected PAPR for I/Q WGN indicate potential non-ideal conditions.  Finally, we presented a real-world example with a spectrogram showing how the mean PAPR formula provides a simple, quick method to check if observations of a given frequency band consist of only thermal noise.

Our theoretical development was based on the assumption of white, uncorrelated I/Q samples.  However, our evaluations of low-pass filtered noise (see Section~\ref{sec:LPFWN}) and a spectrogram with overlapping windows (see Section~\ref{sec:spectrogram}) indicated that the expected mean PAPR is somewhat robust to data correlations.  Specifically, for low-pass filtered noise with digital filter cutoff frequencies greater than $0.25$ cycles per sample, the mean PAPR was very close to the theoretical mean PAPR for uncorrelated white noise.  Additionally, for a measured spectrogram, we found close agreement between the mean PAPR for thermal noise and the theoretical mean PAPR for WGN, even when the spectrogram window overlap was as high as 75\%.  Further investigations, both theoretical and experimental, on the impacts of data correlations on the PAPR distribution for sampled I/Q data is an interesting topic for future work.  

Most of the prior literature on PAPR has focused on using it as a figure of merit for the analog RF transmitter chain in a communication system.  Specifically, there has been a lot of work on PAPR reduction schemes for OFDM communication signals \cite{Rahmatallah2013}   and power amplifiers to accommodate high PAPR \cite{Nikandish2020}.  On the other hand, there are few references that discuss PAPR in the context of digital signal analyzer measurements, with the exception of \cite{Dunsmore2020} and \cite{Keysight2021}.  The present work helps to fill this gap by providing a simple, exact formula for the mean PAPR of WGN in I/Q data and showing how it can be applied to identify non-ideal I/Q acquisitions and to RF channel occupancy assessments.

\section*{Acknowledgments}
The authors would like to thank M. Keith Forsyth, Dazhen Gu, Dan Kuester, Michelle Pirrone, and Todd Schumann for their helpful comments on this article.

\bibliographystyle{IEEEtran}

\bibliography{references}

\end{document}